\documentclass {aastex}

\usepackage{spr-astr-addons}

\begin{document}
\title{The Baltimore and Utrecht models for cluster dissolution}
\author{Henny J. G. L. M. Lamers}
\affil{Astronomical Institute, Utrecht University, Princetonplein
5, 3584 CC Utrecht, The Netherlands, e-mail: \texttt{lamers@astro.uu.nl}}

\begin{abstract}
The analysis of the age distributions of star cluster samples of
different galaxies has resulted in two very different empirical models
for the dissolution of star clusters: the Baltimore model and the
Utrecht model.  I describe these two models and their differences. The
Baltimore model implies that the dissolution of star clusters is mass
independent and that about 90\% of the clusters are destroyed each age
dex, up to an age of about a Gyr, after which point mass-dependent
dissolution from two-body relaxation becomes the dominant mechanism.
In the Utrecht model, cluster dissolution occurs in three stages: (i)
mass-independent infant mortality due to the expulsion of gas up to
about $10^7$ yr; (ii) a phase of slow dynamical evolution with strong
evolutionary fading of the clusters lasting up to about a Gyr; and
(iii) a phase dominated by mass dependent-dissolution, as predicted by
dynamical models. I describe the cluster age distributions for
mass-limited and magnitude-limited cluster samples for both models.  I
refrain from judging the correctness of these models.
\end{abstract}

\section{Introduction}

The dissolution of star clusters in external galaxies can in principle
be derived empirically from the analysis of the age distribution of
detected clusters. If the clusters are destroyed rapidly in their host
galaxy, the number of clusters will decrease rapidly with age, in the
sense that there will be many fewer old than young clusters.  If
cluster destruction is slow, the decrease with age will be more
gentle.

In practice, this method is more complicated than suggested here. This is for several reasons:
\begin{enumerate}
\item One has to assume (or determine) a cluster formation history,
because changes in the cluster formation rate with time will affect
the shape of the age distribution of the observed clusters.
\item Clusters fade with age due to stellar evolution. If the cluster
sample is magnitude limited, older clusters will drop out of the sample
because they are fainter than the detection limit.
\item For both magnitude-limited and mass-limited samples, the samples
have to be corrected for incompleteness. This implies that the study
needs to be restricted to the age and mass range over which reliable
completeness corrections can be done.
\item The accuracy of the age and mass determinations, which are based
on the fitting of the photometric spectral energy distributions with
cluster models, has to be taken into account.
\end{enumerate}

The dissolution of star clusters has been derived from the age
distribution of cluster samples.  These studies have resulted in two
empirical models for cluster dissolution, which I will refer to as the
`Baltimore model' and the `Utrecht model'.

For the interpretation of the age distribution using the Baltimore
model, see Fall et al. (2005), Whitmore et al. (2007; Antennae, solar
neighbourhood, SMC), Chandar et al. (2006; SMC) and Chandar (these
proceedings).  For interpretations based on the Utrecht model, see
Boutloukos \& Lamers (2003, hereafter BL03; solar neighbourhood, SMC,
M33, M51), Gieles et al. (2005a; M51), Lamers et al. (2005a) and
Lamers \& Gieles (2006; solar neighbourhood), Gieles et al. (2007b;
SMC), and Gieles (these proceedings).
 
In this paper, I will explain the differences between two models. I
will try to give an unbiased description of the two models and refrain
from judging the correctness of the models. This implies that the
reader may find different interpretations of the same data sets.

The age distributions of both magnitude-limited samples and
mass-limited cluster samples are compared in Fig. 1, and described
below.  The main observable difference between the two interpretations
is the presence (Utrecht) or absence (Baltimore) of a bend in the age
distributions, and the presence (Utrecht) or absence (Baltimore) of a
flat part in the age distribution of mass-limited cluster samples.

\begin{figure}[!h]
\plotone{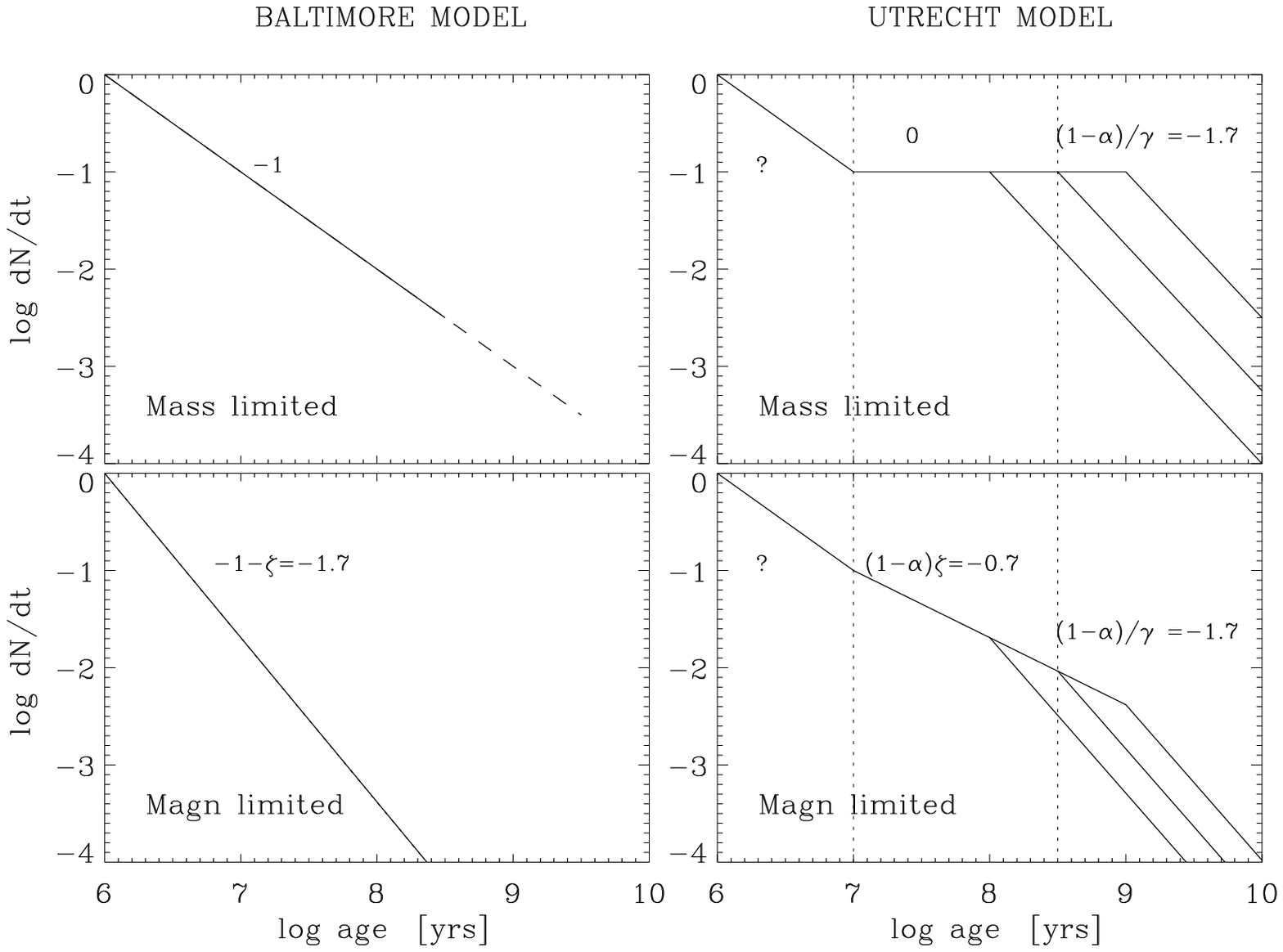}
\caption[]{Schematic representation of the age distributions of
cluster samples which are mass limited (upper panels) or magnitude
limited (lower panels) for the Baltimore and Utrecht models for
cluster dissolution. \\In the Baltimore model cluster dissolution is
mass independent, and removes 90\% of the clusters each age dex up to
about 1 Gyr.  The dashed line indicates that the Baltimore group finds
it impossible to extend the analysis of their data sets beyound this
point, where they assume that two-body relaxation becomes
important.\\The Utrecht dissolution model occurs in three phases: (i)
mass-independent infant mortality, (ii) slow dynamical evolution with
evolutionary fading, and (iii) mass-dependent dissolution. The
transition between (ii) and (iii) depends on the conditions in the
host galaxy. In an environment with a strong tidal field or a high
density of GMCs, the transition will occur at younger ages than in a
quiescent galaxy.\\The slopes of the relations are indicated. The
parameters are $\alpha \simeq 2$ for the cluster initial mass
function, $\zeta \simeq 0.6$ to 1 for evolutionary fading, and $\gamma
\simeq 0.6$ for mass-dependent dissolution of the form $t_{\rm dis}
\propto M^{-\gamma}$.  (The predicted bends of the Utrecht model will
be less sharp and more gentle than shown in this systematic picture.)
See text for explanation} \label{fig1}
\end{figure}

\section{The Baltimore model for cluster dissolution}

The Baltimore model for cluster dissolution is based on the observed
age distribution of cluster samples in the Antennae, the solar
neighbourhood, the SMC and the LMC (Whitmore et al. 2007; Chandar et
al. 2006; Chandar, these proceedings).

In this model, the age distribution of a {\bf mass-limited cluster
sample} is a straight line with a slope of $-1$ in a $\log({\rm
d}N/{\rm d}t)$ versus log(age) plot, where ${\rm d}N/{\rm d}t$ is the
number of clusters in a logarithmic age bin divided by the linear age
range of that bin. The same slope is also found when different mass
cuts are applied to the data sets. This implies that dissolution is
independent of the mass of the clusters up to approximately 1 Gyr, at
which point incompleteness limits set in.  I will refer to this as
mass-independent dissolution, MID\footnote{Mass-independent
dissolution is sometimes called `infant mortality'. However, I suggest
that this name is reserved for the dissolution of star clusters due to
the gas expulsion during the first $10^6$ to $10^7$ yr, as originally
proposed by Lada \& Lada (2003).}.  The steep slope, i.e., $-1$, of
the logarithmic age distribution implies that 90\% of the clusters are
dissolved in each age dex, independent of their mass.

For {\bf magnitude-limited cluster samples}, the slope of the age
distribution must be steeper.  This is because both cluster
dissolution and evolutionary fading remove clusters from a sample
above a certain brightness limit. The fading of the clusters due to
stellar evolution is described by cluster evolution models, such as
the models of Bruzual \& Charlot (2003), the {\sc Starburst99} models
(Leitherer et al. 1999), the {\sc galev} models (Anders et al. 2003),
etc.  After the first $\sim$10 Myr, the fading of clusters of constant
mass can be approximated by $L_{\lambda} \propto t^{-\zeta}$ with
$\zeta \simeq 0.69$, 0.85 and 1.0 for the $V, B$, and $U$ filters,
respectively. This implies that the predicted age distribution of the
Baltimore model for the magnitude-limited samples will be steep, with
a slope of $-1-\zeta$, i.e., between about $-1.7$ and $-2$. (The
Baltimore group has thus far not studied the age distribution of
magnitude-limited samples.)

\section{The Utrecht model for cluster dissolution}

The Utrecht model for cluster dissolution is based on the analysis of
magnitude-limited cluster samples of the solar neighbourhood, the SMC,
M33 and M51, originally started by Boutloukos \& Lamers (2003,
hereafter BL03).  In this first paper, cluster dissolution was assumed
to be instantaneous, but the analysis has been refined for gradual
cluster dissolution as described by Lamers et al. (2005).

In this model, the dissolution of clusters occurs in three steps, in
agreement with some predictions based on the dynamical evolution of
clusters.
\begin{enumerate}
\item During the first 10 to 20 Myr, clusters dissolve due to the
expulsion of the remaining gas.  This dissolution is assumed to be
independent of mass (e.g. Bastian \& Goodwin 2006).  The fraction of
the clusters that dissolve at such an early phase is called the
`infant mortality rate' which may be as high as 30 to 90\% (Lada \&
Lada, 2003; Fall et al. 2005; Bastian et al. 2005; Lamers \& Gieles
2006).

\item During the next $10^8$ to $10^9$ yr, the number of observable
clusters in a magnitude-limited sample decreases due to evolutionary
fading. The slope of the logarithmic ${\rm d}N/{\rm d}t$ distribution
will be about $\zeta(1-\alpha)$ or about $-1$ to $-0.6$, where
$-\alpha \simeq -2$ is the slope of the initial cluster mass function
(BL03).

\item At ages older than about a Gyr, but dependent on the local
conditions in the host galaxy, cluster dissolution by tidal effects
due to the galactic tidal field and due to shocks by spiral arms and
passing giant molecular clouds becomes important. These dissolution
effects are mass dependent, approximately as $t_{\rm dis} =
M^{-\gamma}$, with $\gamma \simeq 0.6$ (BL03; Baumgardt \& Makino
2003; Gieles et al. 2006b, 2007). During this phase, low-mass clusters
are more easily destroyed than massive clusters.
\end{enumerate}

The resulting age distribution of {\bf magnitude-limi\-ted cluster
samples} correspondingly consists of three parts: (i) the steep
decrease due to infant mortality (the shape of this decrease is not
known), (ii) a slower decrease between $10^7$ and about $10^9$ yr with
a slope of $(1-\alpha)\zeta \simeq -0.6$ to $-1$ due to evolutionary
fading, and (iii) a steeper slope of $(1-\alpha)/\gamma \simeq -1.7$
due to dynamical effects (BL03).  The onset of this decrease depends
on the local conditions. In an environment with a strong tidal field
or a high density of giant molecular clouds, dynamical dissolution is
faster than in more quiescent environments, and so the transition
between regions (ii) and (iii) will be at a younger age.

The age distribution of {\bf mass-limited cluster samples} will also
show the three regimes. (i) The slope due to infant mortality will be
the same as for magnitude-limited samples if infant mortality is mass
independent; (ii) evolutionary fading does not affect the age
distribution of mass-limited samples, so the age distribution will be
flat for mass-limited samples; and (iii) in the last phase, where
dynamical dissolution is more important than evolutionary fading, the
age distribution will show a steep slope of approximately
$(1-\alpha)/\gamma \simeq -1.7$ (BL03).

Gieles (these proceedings) has shown that the age distribution of the
brightest clusters in a galaxy can also be used to derive information
about the dissolution.

\acknowledgments I thank the organizers of this workshop, Enrique
Perez and Richard de Grijs, for the opportunity to explain the
difference between the two models.  Brad Whitmore is gratefully
acknowledged for his constructive comments.

%
%

\end{document}